\documentclass[pre,twocolumn,showpacs,preprintnumbers,amsmath,amssymb]{revtex4} %twocolumn

\usepackage{graphicx}% Include figure files
\usepackage{dcolumn}% Align table columns on decimal point
\usepackage{bm}% bold math
\begin{document}
\title{Analytical Investigation of Innovation Dynamics Considering 
Stochasticity in the Evaluation of Fitness} 
\author{Dirk Helbing, Martin Treiber}
\affiliation{Dresden University of Technology, Andreas-Schubert-Str. 23, 01062 Dresden, Germany}
\author{and Nicole J. Saam} 
\affiliation{Ludwig Maximilian University Munich, Konradstr. 6,  80801 Munich, Germany}

\begin{abstract}
We investigate a selection-mutation model for the dynamics of technological innovation,
a special case of reaction-diffusion equations.
Although mutations are assumed to increase the variety of technologies, not their average success
(``fitness''), they are an essential prerequisite for innovation. Together with
a selection of above-average technologies due to imitation behavior, 
they are the ``driving force'' for the continuous
increase in fitness. We will give analytical solutions 
for the probability distribution of technologies for special cases and in the limit 
of large times. 
\par
The selection dynamics is modelled by a ``proportional imitation'' of 
better technologies. However, the assessment of a technology's fitness may
be imperfect and, therefore, vary stochastically. 
We will derive conditions, under which wrong assessment of fitness can
accelerate the innovation dynamics, as it has been found in some surprising numerical
investigations.
\end{abstract}
%Pattern selection; pattern formation
\maketitle
%\draft

{\em Introduction.}
Innovation dynamics has not only been studied by economists and social
scientists (see Refs. in \cite{Hub}), but also recently attracted an 
increasing interest by physicists \cite{Hub,Llas,Guard}. Ebeling {\em et al.}, for example,
investigate the competition of an innovation with already 
established technologies \cite{BrucEbJiMo96,Ebeling}. 
\par
In a wider perspective,
the selection-mutuation equations discussed in this manuscript are a special case 
of reaction-diffusion equations, where technological imitation processes are analogous
to chemical reactions and diffusion originates from mutation processes. Due to their
non-linearity, reaction-mutation equations are known to display interesting pattern-formation phenomena
such as (spiral) waves \cite{react,turing1} 
or Turing patterns \cite{turing1,turing2}. However, this subject will not be the
focus of this paper. Instead, we will try to understand some surprising observations in
a microsimulation model of innovation dynamics by Saam \cite{Saam}.
\par
Our innovation model  assumes that the success of a company's technology can be
expressed by a ``fitness'' value $x$. The technologies of superior companies are
imitated (copied) with a certain probability per unit time specified later.
Moreover, all companies perform innovative activities, which are
modeled by random mutations of the currently applied technologies. Assuming 
a Brownian motion for the mutation of fitness values, 
they change from some value $x$ at time $t$ to some other value
$z$ at time $t+\Delta t$ with a Gaussian transition probability 
\begin{equation}
 p(z(t+\Delta t)|x(t))  = 
 \frac{\mbox{e}^{-[z(t+\Delta t)-x(t)]^2/(2\theta\nu \, \Delta t)} }
 {\sqrt{2\pi \theta \nu \, \Delta t}}  \, .  
\label{asin}
\end{equation}
Here, $\nu$ is the frequency, $\nu \, \Delta t$ the number and $\sqrt{\theta}$ the standard deviation
(average size) of mutations. According to Eq.~(\ref{asin}), mutations may
be beneficial, but they may also decrease the fitness of technologies. Therefore, the 
observed increase in the average fitness would have to be a consequence of the assumed
imitation of superior technologies. 
\par 
Patents on the leading technologies could suppress such imitation and, therefore, potentially reduce the
speed of innovation, i.e. the increase in the average fitness of technologies. However, under 
certain conditions, a misperception of fitness values \cite{Hub} seems to neutralize or even over-compensate
such a deceleration effect \cite{Saam}. Our paper will try to achieve an analytical understanding
of these interesting findings.

{\em Formulation of the Innovation Model.} 
For the sake of analytical treatment, 
let us assume a continuous spectrum of possible fitness values $x$.
Moreover, let $P(x,t)\, dx$ represent the probability with which the applied technology of a company
has a fitness value between $x$ and $x + dx$ at time $t$. In addition, we will assume that companies 
imitate technologies with a higher fitness $z > x$
and that the imitation rate increases proportionally with the related increase $z - x$ in success.
If $z - x$ is negative, the imitation rate should be zero. Then, given
the presently applied technology has the fitness $x$, the 
respective imitation rate $w_1(z|x,t)$ to copy another technology of fitness $z$, is 
\cite{game,kluwer} 
\begin{equation}
 w_1(z|x,t) = \lambda P(z,t) \max( z-x,0 ) \, , 
\label{ueber}
\end{equation}
as the imitation rate is proportional to the occurence probability density $P(z,t)$ of the imitated technology $z$.
The parameter $\lambda > 0$ has the meaning of an imitation frequency.
\par
In order to calculate the change of the occurence probability density $P(x,t)$ due to imitation processes, we have
to insert $w(z|x,t) = w_1(z|x,t)$ 
into the continuous master equation \cite{kluwer,weidlich,evolbook}
\begin{equation}
 \frac{dP(x,t)}{dt} = \int dz \, [w(x|z,t)P(z,t) - w(z|x,t)P(x,t)] \, ,
\end{equation}
which leads to 
\begin{eqnarray}
\frac{dP(x,t)}{dt} &=& \lambda \int dz \, P(x,t) P(z,t) \nonumber \\ 
 & & \quad \times \underbrace{[ \max(x -z ,0) -  \max( z - x,0 ) ]}_{= x-z} \\
&=& \lambda P(x,t) \bigg[ x - \!\!\!\!\underbrace{\int dz \,  z P(z,t)}_{\mbox{\footnotesize\rm 
=~av. fitness $\langle x \rangle_t$}} \!\!\!\bigg] \, ,
\label{selek}
\end{eqnarray}
because of the normalization condition
$\int dz \, P(z,t) = 1$.
Equation (\ref{selek}) is related to the selection equation from evolutionary
biology \cite{eigen,evolbook} and to the game-dynamical equations \cite{HoSi,game,kluwer}.
In order to take into account new inventions, with
\begin{equation}
 w(z|x,t) = \lim_{\Delta t \rightarrow 0} \frac{p(z(t+\Delta t)|x(t))}{\Delta t} 
+ w_1(z|x,t) 
\end{equation}
we will additionally assume random transitions to other technologies $z$ with the rate
$w_0(z|x,t)= w(z|x,t)- w_1(z|x,t)$. According to this,
we simply treat inventions as unbiased random mutations, which may either increase or
decrease the fitness of the pursued technology, as in complex (technological) systems
it is sometimes hard to judge before, whether some change will cause an improvement or not.
Although the assumption of unbiasedness may be questioned, it is favorable to gain 
non-trivial conclusions. The resulting equation for the temporal change
of $P(x,t)$ can be approximated by the following Fokker-Planck type of equation \cite{Risken}:
\begin{equation}
  \frac{\partial P(x,t)}{\partial t}  = \lambda P(x,t) \bigg[ x - 
 \underbrace{\int dz \, z P(z,t)}_{=\langle x \rangle_t} \bigg] 
 + D \frac{\partial^2 P(x,t)}{\partial x^2} \, .
\label{diff}
\end{equation}
The quantity $D = \nu \theta /2$ has the meaning of a diffusion coefficient.  For the mean fitness
$\langle x \rangle_t$ one can derive the differential equation \cite{evolbook,kluwer}
\begin{equation}
 \frac{d\langle x\rangle_t}{dt} = \lambda \mbox{Var}_t(x) \, ,
\label{variety}
\end{equation}
while for the variance $\mbox{Var}_t(x) = \langle (x - \langle x \rangle_t)^2 \rangle_t$, we
find 
\begin{equation}
\frac{d\mbox{Var}_t(x)}{dt} =  \lambda \langle (x - \langle x \rangle_t)^3 \rangle_t + 2D \, .
\label{two}
\end{equation}
The term $\langle (x - \langle x \rangle_t)^3 \rangle_t$ disappears if the
distribution $P(x,t)$ has a vanishing skewness $\gamma$, e.g. for a Gaussian distribution.
Equation (\ref{variety}) shows that the diversity Var$_t(x)$ is a key factor for a high 
innovation speed $d\langle x \rangle_t/dt$.
\par
\begin{figure}[htbp]
\begin{center}
\vspace*{-5mm}
\includegraphics[width=9cm]{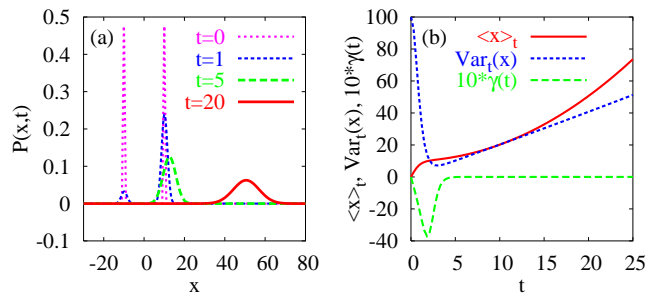}
\vspace*{-14mm}
\end{center}
\caption[]{(Color Online)(a) Computer simulation of Eq.~(\ref{diff}) with $D=1$ and $\lambda=0.1$.
Starting with a two-peak initial distribution $P(x,0)$, one can see that the superposition
law does not apply due to the non-linearity of our innovation model. Nevertheless, after some time
($t \ge 5$) we find the Gaussian distribution (\ref{app}). 
(b) After the initially developing skewness $\gamma(t) 
= \langle (x - \langle x \rangle_t)^3\rangle_t/ \mbox{Var}_t(x)^{3/2}$ has vanished ($t \ge 5$),
the variance Var$_t(x)$ grows linearly and the mean fitness $\langle x \rangle_t$ quadratically,
in accordance with Eqs.~(\ref{lin}) and (\ref{quad}).\label{Fig1}} 
\end{figure}

Equation (\ref{diff}) can be viewed as a special case of reaction-diffusion equations \cite{react,turing1}, 
where the diffusion term originates from mutations and the non-linear
reaction term from imitative pair interactions \cite{game,kluwer}. 
It is known to have a complicated formal solution, which is uniquely 
determined by the initial condition \cite{evolbook,kluwer}. However, this does not help us, 
here. We are rather looking for a qualitative understanding of the dynamic solution 
and, where possible, for explicit analytical expressions. 
In the limiting case $\nu = 0 = D$ of no mutation and no diffusion, all companies will
eventually imitate the technology with the highest fitness $x_{\rm max}$, i.e.
$\lim_{t\rightarrow \infty}P(x,t) = \delta(x - x_{\rm max})$, where $\delta(x)$ denotes 
Dirac's delta function. In the limiting case $\lambda = 0$
of no imitation, Eq.~(\ref{diff}) is just a linear diffusion equation, and its solution is a superposition of
Gaussian distributions with a linearly-in-time increasing variance $2Dt$:
\begin{equation}
 P(x,t) = \int dz \frac{P(z,0)}{\sqrt{4\pi Dt}} \, \mbox{e}^{-(x-z)^2/(4Dt)} \, .
\label{simi}
\end{equation}
If we initially have a two-point distribution $P(x,0)
= p_1 \delta (x - x_1) + p_2 \delta (x-x_2)$ with $p_1+p_2 = 1$, 
both peaks will become broader due to diffusion similarly to Eq.~(\ref{simi}). At the same time,
however, imitation tends to favour the peak around the superior technology's fitness, 
while the peak around the inferior technology's fitness will eventually disappear.
After some time, one finds a unimodal, almost Gaussian distribution, if
$\lambda$ and $D$ are constant, see Fig.~\ref{Fig1}. 
\par
Therefore, let us investigate how a Gaussian distribution develops in time. 
We will show that the non-linear Eq.~(\ref{diff}) has, then, 
a special solution of the form
\begin{equation}
 P(x,t) = \frac{1}{\sqrt{2\pi g(t)}} \, \mbox{e}^{-[x-f(t)]^2/[2g(t)]} \, ,
\label{app}
\end{equation}
where $f(t)$ denotes the mean value and $g(t)$ the variance of the fitness values $x$.
Applying the product, quotient, and chain rules of differential calculus, we find
\begin{eqnarray}
\frac{\partial P(x,t)}{\partial x}&=&-\frac{2[x-f(t)]}{2g(t)}P(x,t) \\
\frac{\partial^{2}P(x,t)}{\partial x^{2}}&=&-\frac{1}{g(t)}P(x,t)+\left\{-\frac{2[x-f(t)]}{2g(t)} \right\}^{2}P(x,t) \nonumber \\
&=&-\frac{1}{g(t)}P(x,t)+\frac{[x-f(t)]^{2}}{g(t)^{2}}P(x,t) \\
\frac{\partial P(x,t)}{\partial t}&=&-\frac{1}{2}\frac{\frac{dg}{dt}}{g(t)}P(x,t)+\frac{[x-f(t)] \frac{df}{dt}}{g(t)}P(x,t) \nonumber \\
&+&\frac{1}{2}\frac{\frac{dg}{dt} [x-f(t)]^{2}}{g(t)^{2}}P(x,t) \, .
\end{eqnarray}
Comparing the expression for $\partial P(x,t)/\partial t$ with the one for
$\partial^2 P(x,t)/\partial x^2$ shows that Eq.~(\ref{app}) implies
\begin{equation}
 \frac{\partial P(x,t)}{\partial t}=\frac{\frac{dg(t)}{dt}}{2}\frac{\partial^{2}P(x,t)}{\partial x^{2}}
 +\frac{[x-f(t)]\frac{df(t)}{dt}}{g(t)}P(x,t) \, .
\label{msg}
\end{equation}
Considering $\langle (x - \langle x \rangle_t)^3 \rangle_t = 0$, $f(t) = \langle x \rangle_t$,
$g(t) = \mbox{Var}_t(x)$, and Eqs.~(\ref{variety}), (\ref{two}),
this partial differential equation indeed agrees with Eq.~(\ref{diff}).
Therefore, Eq.~(\ref{app}) is the unique solution if the initial condition is a Gaussian distribution.
The existence of such a generalized diffusion solution 
is quite surprising in view of the non-linearity of Eq.~(\ref{diff}). This reminds one of the exact solution
of the non-linear Burgers equation \cite{Burgers}.
\par
In the previously assumed case $D = \mbox{const.}$ of constant mutation activity, we obtain 
\begin{eqnarray}
 g(t) &=& g(0) + 2Dt \, , \label{lin} \\
 f(t) &=& f(0) + \lambda g(0) t + \lambda D t^2 \, . 
\label{quad}
\end{eqnarray}
However, for $D(t) = C_0 g(t)$, i.e. if the overall mutation activity would be proportional to
the variety of existing technologies, we would find an exponential growth 
\begin{eqnarray}
 g(t) &=& g(0) \mbox{e}^{2C_0 t} \, , \\
 f(t) &=& \left( f(0) - \frac{\lambda g(0)}{2C_0} \right) + \frac{\lambda g(0)}{2C_0} \mbox{e}^{2C_0t} \, .
\end{eqnarray}
Finally, for $D(t) = C_1 f(t)$, i.e. if the overall mutation activity were proportional
to the average fitness (and the related profits),
we would expect 
\begin{equation}
 f(t) = f(0) \mbox{cosh} (\sqrt{2\lambda C_1}t) 
+  \frac{\sqrt{\lambda}g(0)}{\sqrt{2C_1}} \, \mbox{sinh} (\sqrt{2\lambda C_1}t) \, . 
\label{decay}
\end{equation}
\par 
In summary, starting with a normal distribution of
technologies $x$, the solution of Eq.~(\ref{diff}) stays normally distributed, but the average 
$f(t)$ and variance $g(t)$ 
depend significantly on the diffusion coefficient $D$ and, hence, on the
mutation activity, i.e. $\nu$ and $\theta$. If the diffusion coefficient 
stays constant, the average fitness grows quadratically in time, while it tends to
grow exponentially, if the diffusion coefficient increases proportionally to the mean value $f(t)$ or  
the variance $g(t)$ of the fitness of technologies. An exponential growth is probably 
the more realistic scenario (see, for example, Moore's law).

{\em Imperfect Evaluation of Fitness.}
Let us now discuss a generalization of the above model, assuming that
the fitness is not assessed exactly. For this, let $\xi$ denote the difference between 
the perceived fitness and the actual fitness  $x$ of a technology,  and $z+\zeta$ the perceived
fitness of a technology with fitness $z$. In this case, we have the formula
\begin{equation}
 w_1(z|x,t) = \lambda \max( z + \zeta - x - \xi,0 ) P(z,t) 
\label{ueber2}
\end{equation}
and study the probability density $P'(x,\xi,t) = P_x(\xi) P(x,t)$,
where $P_x(\xi)$ denotes the conditional probability $P(\xi|x)$ that the error in the
fitness estimate is $\xi$, given the actual fitness is $x$. The following
normalization relation applies: $\int d\xi \,  P_x(\xi) = 1$.
Therefore, the resulting differential equation for the distribution $P(x,t)$ of fitness values
becomes
\begin{eqnarray}
 & & \hspace*{-1.2cm} \frac{\partial P(x,t)}{\partial t} \underbrace{\int d\xi \, P_x(\xi)}_{=1} \nonumber \\ 
 &=& \lambda \int dz \int d\xi \int d\zeta \,
 P_x(\xi) P(x,t) P_z(\zeta) P(z,t) \nonumber \\
 & & \times (x + \xi - z - \zeta)  + D \frac{\partial^2 P(x,t)}{\partial x^2} \underbrace{\int d\xi \, P_x(\xi)}_{=1} .\quad
\end{eqnarray}
\par
The solution of this equation depends crucially on the distribution $P_x(\xi)$
and can, in general, not be analytically determined. We will, therefore, restrict to
deriving a generalization of Eq.~(\ref{variety}) in order to see, how imperfect evaluation of a
technology's fitness affects the temporal evolution of the average fitness, if at all.
Applying the method of partial integration to the diffusion term twice and
defining the mean value of a function $G(x,\xi,z,\zeta,t)$ by 
\begin{eqnarray}
 \langle G \rangle_t &=& \int dx \int d\xi \int dz \int d\zeta \, G(x,\xi,z,\zeta,t) \nonumber \\
& & \quad \times P(x,t) P_x(\xi) P(z,t) P_z(\zeta)   \, ,   
\end{eqnarray}
we obtain
\begin{eqnarray}
\frac{d\langle x \rangle_t}{dt} &=& \int dx \, x \frac{\partial P(x,t)}{\partial t} \\
&=& \lambda \int dx \int d\xi \int dz \int d\zeta \, P(x,t) P_x(\xi) P(z,t) \nonumber \\
& & \quad \times P_z(\zeta) ( x^2 + x \xi - x z - x\zeta ) \nonumber \\
&+& D \int dx \, x \frac{\partial^2 P(x,t)}{\partial x^2} \\
&=& \lambda \left[ \langle x^2 \rangle_t + \langle x \xi \rangle_t
- \langle x z \rangle_t - \langle x \zeta \rangle_t \right] 
\nonumber \\
&+&  D \int dx \, \underbrace{\frac{d^2 x}{dx^2}}_{=0}  P(x,t) \, .
\end{eqnarray}
With $\langle z \rangle_t = \langle x \rangle_t$, $\langle \xi \rangle_t = 
\langle \zeta \rangle_t$,
$\langle x z \rangle_t = \langle x\rangle_t \langle z \rangle_t$, Cov$_t(x,\zeta)=0$, 
Cov$_t(x,\xi) = \langle x \xi \rangle_t - \langle x \rangle_t \langle \xi \rangle_t$,
and analogous relationships, we find
\begin{equation}
\frac{d\langle x \rangle_t}{dt} = \lambda \left[ \mbox{Var}_t(x) + \mbox{Cov}_t(x,\xi) \right] \, .
\label{erg}
\end{equation}
\par
Compared to Eq.~(\ref{variety}), imperfect perception of fitness results in the additional
term Cov$_t(x,\xi)$. Accordingly, misperception does not have any influence on the dynamics,
if the deviations $\xi$ are statistically independent of $x$ or $P_x(\xi)$ is not a function of
$x$. If  Cov$_t(x,\xi)<0$, i.e. if leading technologies are systematically underestimated, while
inferior technologies are overestimated, misperception slows down the increase of the average fitness.
However, misperception will speed up the evolution of better technologies,
if Cov$_t(x,\xi)>0$, i.e. if leading technologies are overestimated. This is actually the case in
the simulation model by Saam \cite{Saam}, as it assumes that 
only the technology of the firm with the {\em best}
perceived fitness is copied, while other firms are never imitated, even if their technologies are better
than the own one. The same simulation study has shown that an overestimation 
of the {\em apparently} best technology can even compensate for
inhibitory effects of patents, which were assumed to suppress  the imitation of the {\em actually} 
fittest technology \cite{Saam}.

{\em Summary.}
In this paper, we have analytically investigated the dynamics of a selection-mutation model
for technological innovations, which can be viewed as a special reaction-diffusion model,
but without emergent pattern formation based on a Turing 
instability \cite{turing1,turing2}.
In our case, the imitation of better technologies plays a role analogous to chemical reactions.
Imitation has an effect comparable to the selection of above-average strategies in similar
equations of biological evolution. Inventions have been modeled as unbiased, random
mutations. They may improve the fitness of technologies or deteriorate them. As a consequence,
the observed increase of the average fitness requires a selection process. 
However, a selection process alone also does not cause a persistent increase in the average fitness:
If the mutation frequency $\nu$ and the diffusion coefficient $D$ are 0, 
the fitness of all technologies converge to the highest initial fitness and stay there.
Only if both, random mutations of technologies
and imitation are combined, we have a steady growth of the average fitness, so that after some time 
even a bad technology will be replaced by another one which is much better than
the initially leading technology. 
Interestingly, the speed of the average increase in fitness is proportional
to the variance in the fitness of technologies. 
Therefore, copying other companies' technologies alone does not support a persistent innovation
trend. Instead, diversity is the ``motor'' or ``driving force'' of innovation. 
\par
Finally, we have investigated the effect of misperception of the fitness of technologies. 
It turned out that misperception can be neutral, but it can also speed up or slow
down technological evolution. Our results indicated
that the average fitness will grow faster, if the leading technologies are 
systematically overestimated and the fitness of inferior technologies is underestimated. 
Therefore, excitement for new technologies can speed up innovations even without higher
investments, just because of a bias in the perception of fitness, i.e. a
bias in the imitation behavior of superior technologies.

\end{document}